\documentclass[prl,showpacs,twocolumn,showkeys,amssymb,amsmath,aps]{revtex4}

\usepackage{graphicx}
\usepackage{bm}
\begin{document}
\title{Tsallis statistics generalization of non-equilibrium work relations}
\author{M. Ponmurugan} 
\affiliation{The Institute of Mathematical Sciences, 
C.I.T. Campus, Taramani, Chennai 600113, India}
\date{\today}

\begin{abstract}
We use third constraint formulation of Tsallis statistics and   
derive the $q$-statistics generalization of  
non-equilibrium work relations such as the 
Jarzynski equality and the Crooks fluctuation theorem   
which relate the free energy differences between two equilibrium 
states and the work distribution of the non-equilibrium processes.
 
\end{abstract}
\pacs{05.70.Ln,05.20.-y,05.70.-a,05.40.-a}
\keywords{Fluctuation Theorem, Jarzynski equality, Tsallis statistics,
non-equilibrium process}
\maketitle

{\it Introduction--}
Recent advances in theory of non-equilibrium 
statistical mechanics have established the methods to 
calculate the free energy differences
between the two equilibrium states of the driven 
system from the non-equilibrium 
work measurements \cite{jar,crooks,col}.  
These methods are generally
called as non-equilibrium work relations and in particular
named as the Jarzynski equality \cite{jar} and the Crooks 
work fluctuation theorem \cite{crooks}.
Consider a system initially
in equilibrium at temperature (inverse) $\beta=1/kT$
($k$ is the Boltzmann constant), which is externally driven from 
its initial equilibrium state $A$ to final 
equilibrium state $B$ by non-equilibrium process.
Let $P_F[\gamma^F]$ be the
probability of the phase space trajectory $\gamma^F$,
for the system driven between 
the two states in forward direction.
This satisfies the Crooks work 
fluctuation theorem \cite{crooks,horo},
\begin{eqnarray}\label{fluc}
  \frac{P_F[\gamma^F]}{P_R[\gamma^R]} &=& e^{\beta(W-\Delta F)},
\end{eqnarray}
where $W$ is the work performed on the driven system, 
$\Delta F$ is the free energy difference between the 
two equilibrium states and $P_R[\gamma^R]$ is the
probability of the phase space trajectory $\gamma^R$,
for the system driven in the  reversed direction.
This is the direct relation between the work dissipation 
and the ratio of probabilities for
the forward and the reversed trajectories and its 
integrated version is Jarzynski equality \cite{jar,jarres},
\begin{eqnarray}\label{cjar}
\langle \mbox{exp}[-\beta W] \rangle =\mbox{exp}[-\beta \Delta F].
\end{eqnarray}
The average $\langle ... \rangle$ is over 
a statistical ensemble of realizations of a given thermodynamic process.

In Crooks work fluctuation theorem, the  
probabilities of non-equilibrium forward and reversed 
trajectories  are related by taking the initial conditions from the 
Boltzmann-Gibb's (BG) equilibrium distribution.
There are very few studies on work fluctuation theorem relating  
the non-equilibrium forward and reversed 
trajectories taken from other statistical distribution \cite{super,camp}.
In particular, finite bath work fluctuation theorem has been derived 
earlier \cite{camp}, which includes the microcanonical
work fluctuation theorem and the Crooks work fluctuation theorem as
the two limiting cases. This has been obtained by  
considering Tsallis statistics \cite{tsall1} as a
finite heat bath statistics \cite{camp}. However, this theorem
generally contains two temperatures instead of one as observed usually in 
non-equilibrium work relations (Eqs.\ref{fluc} and \ref{cjar}). 
There has not been any generalized connection established 
between Tsallis statistics and non-equilibrium work relations
at a single temperature.
Since the generalized connection should exhibit 
interesting applications in complex systems \cite{tbook,andr}, 
in this paper,  we derive the $q$-statistics generalization of 
Jarzynski equality and the Crooks work fluctuation theorem for the 
classical system driven between two equilibrium states by a 
non-equilibrium  process using  Tsallis statistics.

The theory of Tsallis statistics  based on a   
generalized form of entropy, $S_q$, 
characterized by the index $q \in R$, such that $q=1$ recovers the 
standard theory of Boltzmann and Gibbs.
This generalized (Tsallis) entropy is given by the expression \cite{tsall1} 
\begin{eqnarray}\label{qent}
S_q=k\frac{1-c_q}{(q-1)}
\end{eqnarray}
where $k$ is the positive (Boltzmann) constant and 
\begin{eqnarray}\label{cdef}
c_q=\sum_{i=1}^{w}p_i^q.
\end{eqnarray}
Here, $w$ is the total number of microstates of the system and  
$p_i$ is the probability of the system at microstate $i$. 
In the limit $q \to 1$ one can recover BG entropy
\begin{eqnarray}\label{BGe}
S_{BG}=-k \sum_{i=1}^{w}p_i \ \ell n \ p_i.
\end{eqnarray}
Preserving the standard variational principle, Tsallis
established the canonical 
generalized distributions and its refinements \cite{tsall3,tbook}.
Using Tsallis statistics, Chame and Mello have derived the generalization 
of the fluctuation dissipation theorem \cite{fdt}.
Tsallis nonextensive statistical mechanics is also considered 
as an  approach to non-equilibrium stationary states of small or 
complex systems \cite{tbook}. However, its equilibrium 
formulation remains to be valid for obtaining 
thermodynamic properties of the equilibrium system \cite{tbook,thermo}.
There are four versions of Tsallis statistics \cite{tsall4},
in particular, we use the most widely accepted third constraint 
(escorted probability) \cite{tbook} formulation of Tsallis statistics.

In Tsallis third constraint formulation,
the generalized equilibrium canonical distribution 
at $\beta$ is given by \cite{tsall3}
\begin{eqnarray}\label{pdef}
p_i&=&\frac{1}{Z_q} \left [1-(1-q)
\frac{\beta[\epsilon_i-U_q]}{c_q} \right]^{\frac{1}{1-q}} \\ \nonumber
&\equiv& \frac{\mbox{exp}_q[-\beta(\epsilon_i-U_q)/c_q]}{Z_q},
\end{eqnarray}
where $\epsilon_i$ is the energy of the $i$th microstate, $U_q$ is the 
normalized constrained of internal energy which is given by \cite{tsall3,abe}
\begin{eqnarray}\label{udef}
U_q=\frac{1}{c_q}\sum_{i}^{w} p_i^q \epsilon_i
\end{eqnarray}
and $Z_q$ is the $q$-generalized partition function which 
is given by \cite{tbook} 
\begin{eqnarray}\label{zdef}
Z_q=\sum_{i}^{w} \left [1-(1-q)\frac{\beta
[\epsilon_i-U_q]}{c_q} \right]^{\frac{1}{1-q}}
\end{eqnarray} 
The normalization condition of $p_i$ leads to the
relation \cite{tsall3}
\begin{eqnarray}\label{cdef1}
c_q=Z_q^{1-q}.
\end{eqnarray}
This modified formalism also becomes ordinary
canonical ensemble theory in the limit $q \to 1$ \cite{tbook}
with $c_{q=1} =1$ and 
\begin{eqnarray}\label{zq1}
Z_{q=1}=\mbox{exp}[\beta U] \sum_{i}^{w}
\mbox{exp}\left [-\beta \epsilon_i \right ]
\equiv \mbox{exp} \ [\beta (U-F)], 
\end{eqnarray}
where the internal energy 
\begin{eqnarray}\label{uq1}
U&=& \frac{\mbox{exp}[\beta U]}{Z_{q=1}}  \sum_{i}^{w} \epsilon_i \mbox{exp}\left [-\beta \epsilon_i \right] \\ \nonumber
&=&\frac{\sum_{i}^{w} \epsilon_i\mbox{exp}\left [-\beta \epsilon_i \right]}
{\sum_{i}^{w} \mbox{exp}\left [-\beta \epsilon_i \right ]} 
\end{eqnarray}
and the free energy
\begin{eqnarray}\label{fq1}
F=-kT \ \ell n  \sum_{i}^{w} \mbox{exp}\left [-\beta \epsilon_i \right ].
\end{eqnarray}

Consider a system in an initial macrostate $A$ 
(for example closed system of volume $V_i$) which is  
in  equilibrium at $\beta$. The probability for the system 
in a microscopic phase-space (microstate) $\Gamma_A$ is 
given by \cite{tsall3,abelaw1} 
\begin{eqnarray}\label{pa}
P(\Gamma_A)=\frac{1}{Z_q(A)} \left [1-(1-q)\frac{\beta
[H(\Gamma_A)-U_q(A)]}{c_q(A)} \right]^{\frac{1}{1-q}},
\end{eqnarray}
where $H(\Gamma_A)$ is the Hamiltonian for the 
system in a microstate $\Gamma_A$,
$U_q(A)$ is the internal energy  which is the 
(escorted probability) weighted Hamiltonian eigenvalue \cite{tsall3} 
averaged over all microstate in an initial equilibrium 
state $A$ (see, Eqs. \ref{pdef}, \ref{udef} and \ref{zdef}) and
\begin{eqnarray}
Z_q(A)=\sum_{\Gamma_A} \left [1-(1-q)\frac{\beta
[H(\Gamma_A)-U_q(A)]}{c_q(A)} \right]^{\frac{1}{1-q}}
\end{eqnarray}
with
\begin{eqnarray}
c_q(A)&=&\sum_{\Gamma_A} \left [P(\Gamma_A) \right ]^{q}. 
\end{eqnarray}

Suppose the given system evolves in time  
under Hamiltonian dynamics and reaches a different macrostate $B$ 
which is to be in equilibrium at same $\beta$. 
The probability distribution 
for the system in a microstate $\Gamma_B$ is given by,
\begin{eqnarray}\label{pb}
P(\Gamma_B)=\frac{1}{Z_q(B)} \left [1-(1-q)\frac{\beta
[H(\Gamma_B)-U_q(B)]}{c_q(B)} \right]^{\frac{1}{1-q}},
\end{eqnarray}
where $H(\Gamma_B)$ is the Hamiltonian for the 
system in a microstate $\Gamma_B$, $U_q(B)$, $Z_q(B)$
and $c_q(B)$ have same meaning as above but for the 
macrostate $B$. In order to derive the non-equilibrium 
work relations for a given $\beta$, we formulate the problem  
as follows.

{\it Setup--} 
Consider a classical Hamiltonian system in a macrostate $A$ 
which is initially in equilibrium with reservoir at 
inverse temperature $\beta$. Let $\lambda_t$ be an external protocol
applied in the arbitrary time interval $\tau$ to drive the system from its initial equilibrium state $A$ to another state $B$  at  constant bath temperature $\beta$. It is assumed that the final state $B$ is not necessarily to be 
in equilibrium. However, the system in the state $B$ at the constant 
bath temperature relax towards the equilibrium state $B$ 
for the same $\beta$ without doing any work \cite{jarres}.
Let $H(\Gamma_t,\lambda_t)$ is the Hamiltonian with externally controlled 
time-dependent protocol $\lambda_t$ and the phase-space
coordinates of the system, $\Gamma_t$ at a particular time $t$.
At $t=0$, the system Hamiltonian which is in any one of the 
microstate $\Gamma_A$ is $H(\Gamma_0,\lambda_0)=H(\Gamma_A)$; and 
at time $t=\tau$, the system Hamiltonian is $H(\Gamma_{\tau},\lambda_{\tau})$. 
Let $\gamma$ denote the entire trajectory of the driven system 
from  $t=0$ to $\tau$. One can obtain the statistical ensemble
of possible realizations by performing the above process repeatedly. 
In following the refs \cite{jarres,horo,camp},
the work performed on the system for a given trajectory can be 
defined as
\begin{eqnarray}\label{workdef}
W &=& H(\Gamma_{\tau},\lambda_{\tau})-H(\Gamma_0,\lambda_0) \\ \nonumber
  &\equiv& H(\Gamma_{\tau},\lambda_{\tau})-H(\Gamma_A). 
\end{eqnarray}
It should be noted that the work defined in Eq.(\ref{workdef})
is different from the $q$-dependent work as 
given in ref.\cite{tsall3} (see also refs.\cite{abelaw1,abelaw2}).
Different definition of work and its physical meaning has been 
discussed in detail in refs.\cite{horo,wj1,wc1}.

During the time interval $\tau$ in which the system 
is driven, we assume that the 
reservoir should always be in equilibrium at a given $\beta$ \cite{eqres}. 
In such a case, the total heat transferred by the
system can be written into two part as
\begin{eqnarray}\label{qQsplt}
Q &=& Q_q^d+Q^r,
\end{eqnarray}
where $Q_q^d$ is the $q$-dependent heat transferred
between the system and the reservoir which should
preserve the ($q$-dependent) equilibrium 
nature of the reservoir and  
$Q^r$ is the heat transferred when the system relaxes 
towards the required equilibrium state from the final state $B$ 
at a given $\beta$ \cite{flaw}. 
There is no work performed on the system    
during relaxation \cite{jarres}. We can define the 
heat transfer due to relaxation as 
\begin{eqnarray}\label{qQsplt1}
Q^r &=& H(\Gamma_B)-H(\Gamma_{\tau},\lambda_{\tau}).
\end{eqnarray} 
Since the energy conservation is also valid for 
non-equilibrium process \cite{eqres,vanden,neqbook}, 
the above driven system should obey the principle of 
energy conservation for any microscopic trajectory as 
\begin{eqnarray}\label{econ}
W+Q=\Delta U_q,
\end{eqnarray}
where $\Delta U_q=U_q(B)-U_q(A)$. The principle of  energy 
conservation as given in Eq.(\ref{econ}) for  the 
non-equilibrium process \cite{vanden} is different from the 
$q$-generalization of first law as proposed in refs.\cite{tsall3,abelaw1,abelaw2} for equilibrium system.
Therefore, the $q$-dependent heat transferred  by the driven 
system between the two equilibrium state is
\begin{eqnarray}\label{dQ}
Q_q^d&=&\Delta U_q -W -Q^r  \\ \nonumber
     &=&\Delta U_q-[H({\Gamma_B})-H({\Gamma_A})]
\end{eqnarray}
which depends only on the initial and final system states.
Since $Q_q^d$ is independent of the non-equilibrium trajectories 
of the driven system, we impose the condition for 
the system relaxing towards the required ($q$-dependent) 
equilibrium state at a given $\beta$ as 
\begin{eqnarray}\label{cratio}
c_q(B)=c_q(A)\left [1-(1-q)\frac{\beta Q_q^d}{c_q(A)} \right ].
\end{eqnarray}
Using Eq.(\ref{qent}), the above equation can be written as
\begin{eqnarray}
-\beta Q_q^d &=&  \frac {1-c_q(B)}{q-1}-\frac {1-c_q(A)}{q-1} \\ \nonumber
    &=& \frac{1}{k} [S_q(B)-S_q(A)] \\ \nonumber
    &=& \frac{\sigma_q}{k}
\end{eqnarray}
where $\sigma_q=S_q(B)-S_q(A)$ is the $q$-dependent
change in entropy of the equilibrium system \cite{tbook}.
From the above condition (Eq. \ref{cratio}), the 
$q$-dependent change in (equilibrium) reservoir entropy 
for the driven non-equilibrium  process is obtained as \cite{vanden,prig,espo} 
\begin{eqnarray}\label{slaw}
\frac{\Delta S^r_q}{k}&=& \beta Q_q^d.
\end{eqnarray}
The above equation provides 
the consistent usage of $\beta=1/kT$ as the (inverse)
reservoir  temperature in Tsallis statistics \cite{abelaw2}.

The usage of thermodynamic laws are not mandatory for the proof of 
non-equilibrium work relations \cite{jarres,horo}. 
Since both internal energy, applied work and heat are  
formulated clearly for the above driven system \cite{seifert1},
we use Eq.(\ref{econ}) and Eq.(\ref{cratio}) and 
obtain the $q$-generalized non-equilibrium work relations 
in which the initial probability distributions 
(Eqs. \ref{pa} and \ref{pb}) are taken from the  Tsallis statistics.

{\it $q$-generalized Jarzynski equality--}
In order to obtain the $q$-generalized version of Jarzynski 
equality for the above driven process, one can take the following 
$q$-exponential average over an ensemble of realizations 
in which the initial distribution is taken from  Tsallis statistics.
\begin{eqnarray}
\left <e_q^{\frac{-\beta(W+Q^r-\Delta U_q)}{c_q(B)}}\right> 
=\int e_q^{\frac{-\beta(W+Q^r-\Delta U_q)}{c_q(B)}} P(\Gamma_A) \ d\Gamma_A.
\end{eqnarray}
Using the $q$-exponential identity 
$e_q^x e_q^y = e_q^{[x+y+(1-q)xy]}$ \cite{tbook}
and Eq.(\ref{pa}), the integral of the above equation 
can be rewritten as
\begin{eqnarray}\label{dave}
\left <e_q^{\frac{-\beta(W+Q^r-\Delta U_q)}{c_q(B)}}\right> 
&=& \frac{1}{Z_q(A)} \int e_q^{M_q+N_q} d\Gamma_A, 
\end{eqnarray}
where, 
\begin{eqnarray}
M_q&=&-\frac{\beta[W+Q^r-\Delta U_q]}{c_q(B)}. 
\end{eqnarray}
and 
\begin{eqnarray}
N_q&=&-\frac{\beta[H(\Gamma_A)-U_q(A)]}{c_q(A)} \left[1+(1-q)M_q \right]. 
\end{eqnarray}
Using Eq.(\ref{econ}) and Eq.(\ref{cratio}), one can rewrite  
\begin{eqnarray}
1+(1-q)M_q&=& 1+(1-q)\frac{\beta Q_q^d}{c_q(B)} \\ \nonumber
&=&1+\frac{c_q(A)-c_q(B)}{c_q(B)} \\ \nonumber
&=&\frac{c_q(A)}{c_q(B)}.
\end{eqnarray}
Using Eqs.(\ref{workdef} and \ref{qQsplt1}), one can obtained 
\begin{eqnarray}\label{ssum}
M_q+N_q&=&\frac{-\beta}{c_q(B)}[H(\Gamma_B)-U_q(B)]. 
\end{eqnarray}
Therefore, Eq.(\ref{dave}) becomes,
\begin{eqnarray}
\left <e_q^{\frac{-\beta(W+Q^r-\Delta U_q)}{c_q(B)}}\right> &=&\frac{1}{Z_q(A)}\int e_q^{-\frac{\beta [H(\Gamma_B)-U_q(B)]}{c_q(B)}} d\Gamma_A \nonumber.
\end{eqnarray}
Since Hamiltonian dynamics preserve the phase-space volume, 
$d\Gamma_A=d\Gamma_B$ \cite{jarres,horo}
and using Eq.(\ref{pb}) we can rewrite the above equation as
\begin{eqnarray}\label{genjar}
\left <e_q^{\frac{-\beta(W+Q^r-\Delta U_q)}{c_q(B)}}\right> &=&\frac{Z_q(B)}{Z_q(A)}\int P(\Gamma_B) \ d\Gamma_B \\ \nonumber
&=& \frac{Z_q(B)}{Z_q(A)}.
\end{eqnarray}
We have obtained $q$-generalized version of one of the 
non-equilibrium work relation.
It should be noted that $\beta$ appeared in the
$q$-exponential average  to ensure that the 
temperature of the reservoir  remains same,
however, one does not know anything about system 
temperature during the driven process.
Further, $Q^r$ and $c_q(B)$ in the above relation also 
takes care of the heat exchange due to relaxation of the system 
towards the final equilibrium state \cite{flaw}. 
This may provide the possible physical meaning of the 
above average for the driven non-equilibrium process 
instead of thinking as an adhoc method \cite{jarres,flaw}.

In the limit $q \to 1$, $\mbox{exp}_q(x)=\mbox{exp}(x)$ 
and $c_q=1$ \cite{tbook}, then using Eqs.(\ref{zq1}-\ref{fq1}), 
Eq.(\ref{genjar}) becomes
\begin{eqnarray}\label{jar}
\left <\mbox{exp}\left[-\beta(W+Q^r-\Delta U)\right] \right >
&=&\mbox{exp}\left[\beta(\Delta U-\Delta F)\right] \\ \nonumber
\left <\mbox{exp}\left[-\beta (W+Q^r) \right] \right > &=&
\mbox{exp}\left[-\beta \Delta F\right],
\end{eqnarray}
where $\Delta U=U(B)-U(A)$ is the change in internal energy and 
$\Delta F=F(B)-F(A)$ is the change in equilibrium free 
energy of the BG canonical system. Thus, we can 
obtain more general form of Jarzynski equality which includes 
heat due to system relaxation \cite{flaw} for the BG canonical 
system in the limit $q \to 1$. If $Q_r$ is 
within the measurements/numerical error for work calculation 
in  experiments/simulations, the original Jarzynski equality 
(without the heat term due to relaxation \cite{jarres,flaw}) 
can provide the reliable estimates of the free energy differences.

{\it $q$-generalized Crooks Work fluctuation theorem--} 
In order to derive the $q$-generalized version of the 
Crooks work fluctuation theorem, we proceed with the problem 
analogous to ref.\cite{horo} as follows. Since 
$H(\Gamma_0,\lambda_0)=H(\Gamma_A)$, the 
probability of the phase space trajectory $\gamma^F$,
for the system driven between the two equilibrium states 
obtained from the initial equilibrium distribution 
in the forward direction (A to B) is given as \cite{horo}
\begin{eqnarray}\label{pf1}
P_F[\gamma^F]&=&P^{eq}_A(\Gamma_0^F) \\ \nonumber
&=&\frac{1}{Z_q(A)} 
\left [1-(1-q)\frac{\beta[H(\Gamma_A)-U_q(A)]}{c_q(A)} \right]^{\frac{1}{1-q}}.
\end{eqnarray}
Suppose the system is driven from equilibrium state $B$ to state $A$ using 
the time reversed protocol, $\lambda_t^R=\lambda_{\tau-t}^F$,   
$H(\Gamma_t^R,\lambda_t^R)$ is the Hamiltonian  for the externally 
controlled time-dependent protocol $\lambda_t^R$ and, $\Gamma_t^R$
is the phase-space coordinate of the system  at a particular time $t$.
At $t=0$, the system Hamiltonian which is in any one of the 
microstate $\Gamma_B$ is $H(\Gamma_0^R,\lambda_0^R)=H(\Gamma_B)$; and 
at time $t=\tau$, the system Hamiltonian is $H(\Gamma_{\tau}^R,\lambda_{\tau}^R)$. The probability of the phase space trajectory $\gamma^R$, for the system driven in 
reverse direction obtained from the initial equilibrium distribution
is given as \cite{horo} 
\begin{eqnarray}\label{pf2}
P_R[\gamma^R]&=&P^{eq}_B(\Gamma_0^R) \\ \nonumber
&=&\frac{1}{Z_q(B)} 
\left [1-(1-q)\frac{\beta[H(\Gamma_B)-U_q(B)]}{c_q(B)} \right]^{\frac{1}{1-q}}.
\end{eqnarray}
Using Eq.(\ref{workdef}), Eq (\ref{qQsplt1}) and Eq.(\ref{econ}),
Eq.(\ref{pf1}) can be rewritten as 
\begin{eqnarray}
P_F[\gamma^F]=\frac{1}{Z_q(A)} 
\left [1-(1-q)\frac{\beta
[H(\Gamma_B)-U_q(B)+Q_q^d]}{c_q(A)} \right]^{\frac{1}{1-q}}. \nonumber
\end{eqnarray}

\begin{eqnarray}\label{pa1}
P_F[\gamma^F]&=&\frac{1}{Z_q(A)}  
\left [1-(1-q)\frac{\beta Q_q^d}{c_q(A)} \right]^{\frac{1}{1-q}}  \\ \nonumber
& &\left [1-(1-q)\frac{\beta
[H(\Gamma_B)-U_q(B)]}{c_q(A)[1-(1-q)\frac{\beta Q_q^d}{c_q(A)}]} \right]^{\frac{1}{1-q}}.
\end{eqnarray}
Using Eq.(\ref{cratio}) and Eq.(\ref{pf2}), Eq.(\ref{pa1}) becomes,
\begin{eqnarray}
P_F[\gamma^F]=\frac{Z_q(B)}{Z_q(A)}
\left [1-(1-q)\frac{\beta Q_q^d}{c_q(A)} \right]^{\frac{1}{1-q}} P_R[\gamma^R]. 
\end{eqnarray}
We can get from Eq.(\ref{econ}) that
\begin{eqnarray}\label{fl2}
\frac{P_F[\gamma^F]}{P_R[\gamma^R]}&=&\frac{Z_q(B)}{Z_q(A)}  
\left [1+(1-q)\frac{\beta \left [W+Q^r-\Delta U_q \right]}{c_q(A)} \right]^{\frac{1}{1-q}}. \\ \nonumber
&\equiv& \frac{Z_q(B)}{Z_q(A)}  \
\mbox{exp}_q \left[\frac{\beta \left (W+Q^r-\Delta U_q \right)}{c_q(A)} \right].
\end{eqnarray}
We have obtained the $q$-generalized version of another 
non-equilibrium work relation. 
In the limit $q \to 1$, $\mbox{exp}_q(x)=\mbox{exp}(x)$ 
and $c_q=1$ \cite{tbook}, then using Eqs.(\ref{zq1}-\ref{fq1}), 
Eq.(\ref{fl2}) becomes
\begin{eqnarray}\label{fl4}
\frac{P_F[\gamma^F]}{P_R[\gamma^R]}
&=& \mbox{exp} \left[\beta (W+Q^r-\Delta F)\right].  
\end{eqnarray}
Thus, we can obtain more general form of Crooks work fluctuation 
relation which includes heat due to system relaxation \cite{flaw}
for the BG canonical system in the limit $q \to 1$.

In order to obtain the $q$-generalized version of Jarzynski 
equality from the $q$-generalized work fluctuation relation,
one can take  the following $q$-exponential average over ensemble 
of realization in forward direction as
\begin{eqnarray}
\left <e_q^{\frac{-\beta(W+Q^r-\Delta U_q)}{c_q(B)}}\right> 
=\int e_q^{\frac{-\beta(W+Q^r-\Delta U_q)}{c_q(B)}} 
P_F[\gamma^F] \ d\gamma_F.  
\end{eqnarray}
Using Eq.(\ref{fl2}), the above equation can be rewritten as
\begin{eqnarray}\label{jarav}
\left <e_q^{\frac{-\beta(W+Q^r-\Delta U_q)}{c_q(B)}} \right > 
&=& \frac{Z_q(B)}{Z_q(A)} \ I_q,
\end{eqnarray}
where
\begin{eqnarray}
I_q = \int e_q^{\frac{-\beta(W+Q^r-\Delta U_q)}{c_q(B)}} 
e_q^{\frac{\beta(W+Q^r-\Delta U_q)}{c_q(A)}} P_R[\gamma^R] \ d\gamma_R.
\end{eqnarray} 
Since Hamiltonian dynamics preserve the phase-space volume, 
$d\gamma_F=d\gamma_R$ \cite{jarres,horo}.
Using the $q$-exponential identity 
$e_q^x e_q^y = e_q^{[x+y+(1-q)xy]}$ \cite{tbook}, 
we rewrite the integral of the above equation as 
\begin{eqnarray}
I_q=\int e_q^{\beta(W+Q^r-\Delta U_q)D_q} P_R[\gamma^R] \ d\gamma_R.
\end{eqnarray} 
where
\begin{eqnarray}
D_q=\left[\frac{1}{c_q(A)}-\frac{1}{c_q(B)} 
-(1-q)\frac{\beta(W+Q^r-\Delta U_q)}{c_q(A)c_q(B)} \right].
\end{eqnarray}
Using Eq.(\ref{econ}) and Eq.(\ref{cratio}), the above equation 
becomes,
\begin{eqnarray}
D_q=\left[\frac{1}{c_q(A)}-\frac{1}{c_q(B)}-\frac{[c_q(B)-c_q(A)]}{c_q(A)c_q(B)}\right]=0.
\end{eqnarray}
Since $e_q^0=1$, $I_q=\int P_R[\gamma^R] \ d\gamma_R=1$ and 
Eq.(\ref{jarav}) becomes,
\begin{eqnarray}\label{qjar}
\left <\mbox{exp}_q \left[\frac{-\beta(W+Q^r-\Delta U_q)}{c_q(B)}\right] \right > &=& \frac{Z_q(B)}{Z_q(A)}.
\end{eqnarray}
We have obtained  the $q$-generalized version of Jarzynski equality.

{\it Conclusion--}
We have derived  the  more general form of 
Jarzynski equality and the Crooks work fluctuation theorem
which includes the heat due to system relaxation 
in the framework of Tsallis statistics.
Our general result may resolve the criticism
raised earlier \cite{jarres,flaw} for 
original Jarzynski equality. 
In Tsallis third constraint formulation, one cannot directly
obtain the canonical probability distribution  
because the distribution (Eq.\ref{pdef}) is self referential \cite{tbook,tsall4}.
Since $p_i$ depends upon $c_q$, one should iterate 
Eqs.(\ref{pdef}, \ref{zdef} and \ref{cdef1})
repeatedly until numerical consistency is achieved.
Since Tsallis distribution provides potential application in 
various complex systems, we have utilized the self referential 
nature of the Tsallis distribution and have obtained the
$q$-generalized version of non-equilibrium work relations.

There is a general impression among few of us that Tsallis's formalism 
has nothing to do with equilibrium statistical mechanics. 
If the system relaxes towards the equilibrium, one may 
not rule out the equilibrium formulation of the Tsallis statistics.
Although we have taken the initial distribution as the equilibrium, 
our formulation may also be applicable for non-equilibrium 
stationary state conditions.

{\bf Acknowledgements}
I would like to thank  S.S. Naina Mohammed and R. Chandrashekar 
for the valuable discussions during my learning stage of  
Tsallis statistics. I also thank P. Renugambal for 
proof-reading the manuscript. I thank Prof. M. Campisi 
for drawing my attention to his earlier work \cite{camp}.
My special thanks to Prof. C. Tsallis for his personal 
appreciation of the present work. 
\paragraph{email correspondence: mpn@imsc.res.in}

\end{document}